
\magnification=1200
\hfill
\vskip 1.0cm
\def\dsl{\raise.15ex\hbox{/}\kern-.57em\partial}
\def\Dsl{\,\raise.15ex\hbox{/}\mkern-13.5mu D} 
\def\Asl{\,\raise.15ex\hbox{/}\mkern-13.5mu A} 
\def\Bsl{\,\raise.15ex\hbox{/}\mkern-13.5mu B} 

\def \Ch{{\cal H}}
\def \pe{\perp}
\def \12{{1\over 2}}

\def\dd{\delta^{'} (x - x^{'})}
\def\tt{\delta (x - x^{'})}
\def\ij{\delta_{ij}}
\def\L{{\tilde L}}
\def\ab{\delta_{ab}}
\def\S{{\cal S}}
\def \PL{{\it Phys. Lett.}}
\def \PR{{\it Phys. Rev.}}
\def \CMP{{\it Comm. Math. Phys.}}
\def\kk{{1\over 2}\psi + {3\over 4}\chi}
\def\pp{{\psi + {3\over 2}\chi}}
\font\title=cmr12 at 12pt
\footline={\ifnum\pageno=1\hfill\else\hfill\rm\folio\hfill\fi}
\baselineskip=18pt
\vskip 1.0cm
\centerline{\title THE NO-HAIR CONJECTURE IN 2D DILATON}
\centerline{{\title SUPERGRAVITY}}
\vskip 2.0cm
\centerline{{{\bf J. GAMBOA}\footnote{$^\dagger$}{\it Address after
September 1, 1993:
Fachbereich 7 Physik, Universit\"at Siegen, Germany.}
 and {{\bf Y. GEORGELIN}}} }
\centerline{\it Division de Physique Th\'eorique, Institut de Physique
Nucleaire\footnote{$^\ddagger$}{\it Unite de Recherche des Universit\'es
Paris 11 et Paris 6 associ\'ee au CNRS.},}
\centerline{\it F-91406 Cedex, Orsay, France}
\vskip 1.0cm
{\bf Abstract}. We study two dimensional dilaton gravity and supergravity
following
hamiltonian methods. Firstly, we consider the structure of constraints of
$2D$ dilaton gravity and then the $2D$ dilaton supergravity theory is obtained
taking the
square root of the bosonic constraints. We integrate exactly the equations of
motion in both cases and we show that the solutions of the equation of motion
of $2D$ dilaton supergravity differs from the solutions of $2D$ dilaton gravity
only by boundary conditions on the fermionic variables, i.e. the black holes of
$2D$ dilaton supergravity theory are
exactly the same black holes of $2D$ bosonic dilaton gravity modulo
supersymmetry transformations. This result is the bidimensional analogue
of the no-hair theorem for supergravity.
\vskip 0.5cm
\leftline{IPNO-TH 93/25}
\leftline{June 1993}
\vfill
\eject
\hfill
\vskip 0.5cm

\centerline{\bf I. Introduction}
\vskip 0.25cm
The quantization of the gravitational field is a problem plagued by technical
and conceptual difficulties that has resisted a solution for many years
{\bf [1]}. However in spite of these difficulties some remarkable ideas
have emerged which make one think that the quantized gravitational field
necessarily involves new physics {\bf [2]}.

Among these ideas, the evaporation of black holes (BH) {\bf [3]} is
probably the most important result reached in the last twenty years and for
which there is not a definitive explanation in terms of a true quantum
theory of gravity.

Some time after the theoretical discovery of the evaporation
of BH, Hawking {\bf [4]} argued that the evaporation and creation of BH
necessarily imply a radical change of the quantum mechanical laws and for these
reasons he argued that pure states in quantum gravity could evolve to
mixed states. Despite many efforts made in the last years, it has not been
possible to prove this conjecture and for this reason many people think if
simplifications are not introduced in the theory, probably the Hawking's
conjecture never will be proved.

Recently Callan, Giddings, Harvey and Strominger (CGHS) proposed a simplified
two
dimensional gravity model coupled to a dilaton and conformal matter that
has among its properties to be exactly soluble and to contain as a
particular case the black hole solution {\bf [5]}.
The classical solubility of the model could be an indication that
at the quantum level the solution of some old may be solved in terms of a
completely quantized gravity theory.

However, in spite of the intense research in this area {\bf [6]}, there
still remain open several classical
problems that are important to solve in order to see if the CGHS model
keeps some properties of four dimensional gravity. Among
these properties, a no-hair theorem for $2D$ dilaton gravity has been
proved in {\bf [7]} but a similar result for $2D$ dilaton supergravity
theory, to our knowledge, does not exist.

The purpose of the present research is to construct a dilaton supergravity
model directly from the constraints of $2D$ dilaton gravity and to
investigate the no-hair conjecture for the supergravity case.
Our model of $2D$ supergravity is contructed from an action
originally proposed by Russo and Tseytlin in {\bf [8]} which is clasically
equivalent to the original CGHS action. Our main main result will be
that the solutions of the equation of motion of $2D$ dilaton supergravity
coupled to superconformal matter are exactly the $2D$ dilaton gravity coupled
to conformal matter solutions modulo boundary conditions on the fermionic
variables.

The paper is organized as follows: In the next section
we study $2D$ dilaton gravity without fixing
the gauge and we show the equivalence between these contraints and the
constraints of $2D$ dilaton gravity in the conformal gauge. In section 3,
we construct $N=1, \,\,2D$ dilaton supergravity using hamiltonian methods
and compute completely the algebra of constraints. In section 4 we analyze
the equations of motion and we give the general solution for $2D$ dilaton
(super)gravity coupled to (super)conformal matter. We also prove here the
no-hair theorem for $2D$ dilaton supergravity. Finally, in section 5 we
give the conclusions.
\vfill
\eject
\hfill
\vskip 0.15cm
\centerline{\bf 2. 2D Dilaton Gravity: Hamiltonian Analysis}
\vskip 0.25cm

The model considered by CGHS is described by the following action
$$ S = -{1\over 8} \int_{{\cal M}} d^2x \sqrt{-g} \biggl\{ e^{-2\varphi}
\biggl[ R + 4 g^{\alpha\beta} \partial_\alpha \varphi \partial_\beta \varphi
+ \mu^2 \biggr] + 4\sum_{i=1}^N g^{\alpha \beta}
\partial_\alpha f_i \partial_\beta f_i \biggr\}, \eqno(2.1)$$
where ${\cal M}$ is the manifold on which the theory is defined, $\varphi$ is
the dilaton field, $\mu^2$ the cosmological constant and $f_i$ are $N$ scalar
fields that represents the matter degrees of freedom. Here $g^{\alpha\beta}$
is the two
dimensional metric tensor and $R$ the corresponding scalar curvature.

In order to perform the hamiltonian formulation of (2.1), it is convenient
to transform this action to the form action proposed by Russo and Tseytlin
{\bf [8]} by means the transformations
$$\eqalignno{&\phi = {1\over 4}e^{-2 \varphi}, &(2.2)
\cr & h_{\alpha\beta} = {1\over 4}e^{-2\omega} g_{\alpha\beta}, &(2.3) \cr }$$
where
$$\omega = {1\over 2}(\ln \phi - \phi). \eqno(2.4)$$

Then, the action (2.1) becomes
$$S= S_1 + S_2, \eqno(2.5)$$
where
$$S_1 = -{1\over 2}\int d^2x \sqrt{-h}\left[ h^{\alpha\beta}\partial_\alpha
\phi
\partial_\beta + R\phi +{1\over 4}\mu^2 e^\phi \right], \eqno(2.6)$$
$$S_2 = \int d^2x \sum_{i=1}^N  \sqrt{-h}h^{\alpha\beta}
\partial_\alpha f_i \partial_\beta f_i, \eqno(2.7)$$
being $R$ the scalar curvature computed with the metric $h_{\alpha\beta}$.

The form (2.5) is more convenient because it separately represents the pure
gravity ($S_1$) and matter ($S_2$) sectors. Formally the matter sector
is mathematically equivalent to a string, the hamiltonian formulation
of which is known {\bf [9]}. On other hand, the hamiltonian formulation of pure
$2D$ gravity has been also considered {\bf [10-12]} and, as a consequence,
the constraints associated with (2.5) are
$$\eqalignno{&{\Ch}_\pe = {1\over 2} [ {\phi^{'}}^2 - 4\,\,{(h_{11}\pi^{11})}^2
-
4\,\,(h_{11}\pi^{11})P_\phi -
{{h^{'}_{11}}\over h_{11}}\phi^{'} + 2\phi^{''} +{1\over 4} \mu^2 h_{11}\,\,
e^{\phi}] +
\sum_{i=1}^N {1\over 2}(P^2_i + {f^{'}_i}^2), \cr &
{\Ch}_1 = P_\phi \phi^{'} - 2 h_{11}{\pi^{'}}^{11} - \pi^{11}h^{'}_{11} +
\sum_{i=1}^{N}\,\, P_i {f_i}^{'},  &(2.8) \cr}$$
where $P_{\alpha\beta} = {\delta L\over \delta {\dot h^{\alpha\beta}}}$,
$P_\phi = {\delta L\over \delta {\dot \phi}}$ and $P_i =
{\delta L\over \delta {\dot f_i}}$ are the canonical momenta associated with
$h^{\alpha\beta}, \phi$ and $f_i$ respectively.

Geometrically, the constraints (2.8) are the generators of temporal
and spatial deformations and contain all the dynamics of the theory.
Nevertheless, their structure is still complicated and for
this reason it is convenient
to transform these quantities making a change of variables in order to
simplify their structure.

Following {\bf [10,12]} our first canonical transformation is
$$\pi = \pi^{11}h_{11}, \,\,\,\,\, \chi= \ln h_{11}, \eqno(2.9)$$
where $\pi$ and $\chi$ are canonical variables that satisfy
$$ [ \chi(x), \chi(x^{'})]= 0 = [ \pi(x), \pi(x^{'}) ], $$
 $$[ \chi(x), \pi(x^{'}) ] = \delta (x - x^{'}). \eqno(2.10)$$

Using (2.9), the constraints (2.8) becomes
$$\eqalignno{& {\Ch }_\pe = \12 \biggl[ {\phi^{'}}^2  - 4\pi^2 - 4\pi P_\phi
- \chi^{'} \phi^{'} + 2\phi^{''} + 4\mu^2 e^{\chi + \phi}\biggr]
+ \sum_{i=1}^N \12 (P_i^2 + {f^{'}_i}^2 )  \cr &
{\Ch}_1 = P_\phi {\phi}^{'} + \pi \chi^{'} - 2\pi^{'} + \sum_{i}^N P_i f_i^{'}.
&(2.11) \cr}$$

Our second canonical change of variables consists in diagonalizing the
constraints (2.11); thus we propose
$$\eqalignno{& \psi = \phi - \12 \chi, \,\,\,\,\,\,\,\,\,\, b=P_\phi
\cr &
\chi = \chi, \,\,\,\,\,\,\,\,\,\, P = \pi + \12 P_\phi. &(2.12) \cr}$$

This transformation is also canonical because the quantities $(\psi, b)$
and $(P, \chi)$ satisfy
$$[ \psi(x), \psi(x^{'})] = 0 = [ b(x), b(x^{'})],$$
$$[ \psi(x), b(x^{'})] = \delta (x - x^{'}) =
[ \chi(x) , P(x^{'})]. \eqno(2.13)$$

In consequence, the constraints (2.11) can be written in the following form
$$\eqalignno{& {\Ch}_\pe = \12 \biggl[ {\psi^{'}}^2 + b^2 -4P^2 -
{1\over 4}{\chi^{'}}^2
+ 2\psi^{''} + \chi^{''} + {1\over 4}\mu^2 e^{\pp} \biggr] +
\sum_{i=1}^N \12 (P_i^2 + {f^{'}_i}^2 ),
\cr &
{\Ch}_1 = b \psi^{'} + P \chi^{'} - 2P^{'} + b^{'} + \sum_{i=1}^N P_i {f_i}^{'}
,
&(2.14) \cr}$$

Explicit calculation shows that the algebra of constraints is
$$\eqalignno{& \lbrack \Ch_\pe (x), \Ch_\pe (x^{'}) \rbrack =
( \Ch_1 (x) + \Ch_1 (x^{'}) )\dd,
\cr &
\lbrack \Ch_\pe (x), \Ch_1 (x^{'}) \rbrack = ( \Ch_\pe (x)
+ \Ch_\pe (x^{'} )) \delta (x - x^{'}), &(2.15)
\cr &
\lbrack \Ch_1 (x) , \Ch_1 (x^{'}) \rbrack =
( \Ch_1 (x) + \Ch_1 (x^{'}) \delta (x - x^{'}), \cr}$$
that is, the usual diffeomorphism algebra.

The canonical hamiltonian for this theory is
$$H_c = \int dx \lbrack N^\pe \Ch_\pe + N^1 \Ch_1 \rbrack, \eqno(2.16)$$
where $N^\pe$ and $N^1$ are the Lagrange multipliers
associated with the first
class constraints $\Ch_\pe$ and $\Ch_1$ respectively.

The expression (2.16) is the starting point for computing the invariant ADM
mass
and this (in the context of $2D$ dilaton gravity) has been carried out by
Bilal and Kogan {\bf [13]} and de Alwis in {\bf [14]}.

In analogy with string theory, we can redefine the constraints
(2.14) in the following form
$$\eqalignno{ L_\pm = \Ch_\pe \pm \Ch_1 &=
\12 (h_\pm^2 \pm 2h^{'}_\pm - J^2_\mp \mp 2J^{'}_\mp + {1\over 4}\mu^2
e^{\pp} )
\cr &
+ \sum_{i=1}^N {1\over 2}Q_{\pm,i}^2, &(2.17) \cr}$$
where $h_a$, $J_a$ and $Q_{\pm, i}$ are defined as
$$\eqalignno{ &h_a = b + a\psi^{'},
\cr &
J_a = 2P + {a\over 2}\chi^{'}, &(2.18)
\cr &
Q_{i,a} = P_i + af^{'}_i, \,\,\,\,\,\,\,\,\,\,(a=\pm 1)\cr}$$
and satisfy the algebra
$$\eqalignno{& \lbrack h_a (x) , h_b (x^{'}) \rbrack = 2 a \delta_{ab} \dd,
\cr &
\lbrack J_a (x) , J_b (x^{'}) \rbrack = 2 a \ab \dd, &(2.19)
\cr &
\lbrack Q_{a,i} (x) , Q_{b,j} (x^{'}) \rbrack = 2a \ab \delta_{ij} \dd.
\cr}$$

Using (2.19) the algebra of constraints is now
$$\lbrack L_a (x) , L_b (x^{'}) \rbrack = 2 a \ab (L_a (x) +
L_a (x^{'}) ) \dd. \eqno(2.21)$$

In the compact case, of course, (2.20) define a classical
Virasoro algebra with generators $L_n$. In the non-compact case, however,
the situation is more involved because the Fourier expansion, generally
speaking, is not defined. Even so, some progress has been reported recently
{\bf [15]}.
\vskip 0.50cm
\centerline{\bf 3. 2D Dilaton Supergravity from 2D Dilaton Gravity}
\vskip 0.25cm

In this section we derive $2D$ dilaton gravity directly from the results
obtained in the previous section.

The procedure we follow here is the square root method proposed originally
by Dirac in 1928 {\bf [16]}. His central idea was to construct a fermionic
operator ${\hat \S}$ such that its square root gives the Klein-Gordon equation.
The classical version of this procedure and their relation with
supersymmetry is due to Berezin and Marinov {\bf [17]} and their
procedure can be summarized by the following scheme
$$\eqalignno{& Operator\rightarrow Constraint
\cr &
(Anti)Conmutator \rightarrow (Anti)Symmetric \,\, Poisson \,\, Bracket \cr}$$

In the context of four dimensional supergravity theory, this procedure was
used in {\bf [18]} and is reviewed in {\bf [19]} for the spinning string
case and in {\bf [20]} for the supergravity case.

In order to construct the fermionic constraint, let us start with
the following ansatz {\bf [21]}
$$\eqalignno{ \S_a = &\alpha \Gamma_a h_a + \beta \Gamma^{'}_a +
\gamma \Theta_{-a} J_{-a} + \delta \Theta^{'}_{-a} +
(\omega \Theta_a + \kappa \Gamma_{-a} ) e^{ \12 \psi + {3\over 4}\xi}
\cr &
+ \rho \sum_{i=1}^N \Xi_{i,a} Q_{i,a}, (no\,\,summing\,\,a) &(3.1) \cr}$$
where $\alpha, \beta, \gamma, \delta, \omega, \kappa$ and $\rho$ are unknown
coefficients that will be fixed at the end of the calculation. The spinors
$\Gamma_a, \Theta_a$ and $ \Xi_{i,a}$ are real fermionic variables that,
by definition, satisfy the following Clifford algebra
$$ \eqalignno{&\{ \Gamma_a (x) , \Gamma_b (x^{'}) \} = i
\delta_{ab} \tt,
\cr &
\{ \Theta_a (x) , \Theta_b (x^{'}) \} = -i \delta_{ab} \tt, &(3.2)
\cr &
\{ \Xi_{i,a} (x) , \Xi_{j,b} \} = i\ab \ij \tt. \cr}$$

Following the square root method, the aim now is to compute the symmetric
Poisson bracket
$$\{ \S_a(x), \S_b(x^{'}) \} = i \ab {\tilde L}_a (x) \tt, \eqno(3.3)$$
in order to find the fermionic corrections introduced in the bosonic
constraints $L_a (x)$.

A straightforward calculation gives for ${\tilde L}_a$
$$\eqalignno{ {\tilde L}_a = & \alpha^2 h^2_a + \alpha\beta h^{'}_a
- 2i \alpha^2 a\Gamma_a {\Gamma}^{'}_a - \gamma^2 J^2_{-a}
\cr &
- \gamma \delta J^{'}_{-a} + 2i \gamma^2 a\Theta_{-a} \Theta^{'}_{-a}
+ (\kappa^2 - \omega^2) e^{\pp}
\cr &
+ \rho^2 \sum_{i=1}^N ( Q^2_{a,i} - 2i a\Xi_{a,i} \Xi^{'}_{a,i}). &(3.4) \cr}$$

The coefficients that appear in (3.4) can be explicitly evaluated in the
limit
$$ (\Gamma_a, \Theta_a, \Xi_{i,a}) \rightarrow 0.$$

In fact, after comparing with (2.19) we find\footnote{$^1$}{The reader
might note here that we have chosen the plus sign in front of the
coefficients. This is not a loss of generality, because the same situation
occurs, for instance, when we construct the spinning particle from the
spinless particle.}
$$\alpha = {1\over \beta} = {1\over \sqrt{2}}, \,\,\,\,
\gamma = {1\over \delta} = {1\over \sqrt{2}}\,\,\,\, \rho = {1\over \sqrt{2}},
 \eqno(3.5)$$
$$ \kappa^2 - \omega^2 = {\mu^2\over 8}, \eqno(3.6)$$
of course, the next step is to verify if the constraints ${\tilde L}_a$ and
$\S_a$ satisfy a closed superalgebra.

Computing this algebra, we find
$$\eqalignno{& \{ \S_a (x) , \S_b (x^{'}) \} = i \ab \L_a (x) \tt,
\cr &
[ \L_a (x) , \S_b(x^{'}) ] = \ab ( \S_a (x) + 2 \S_a (x^{'}) ) \dd, &(3.7)
\cr &
[ \L_a (x) , \L_b (x^{'}) ] = \ab ( \L_a (x) + \L_a (x^{'}) ) \dd. \cr}$$

The superalgebra (3.7) deserves two comments: 1)  As in $2D$
dilaton gravity, there are no $\delta^{'''}(x - x^{'})$ terms present in
the superalgebra and the total central charge for this model is zero; 2)
Our model depends on an arbitrary contant $\kappa$ (or $omega$) and,
as a consequence, in this dilaton $2D$ supergravity theory the cosmological
constant
can be positive or negative. It is interesting to note that a negative
cosmological constant is mandatory in a theory with only scalar field, such
as a pure Liouville theory.

The constraints $(\L_a, \S_a)$ are first class and the canonical
hamiltonian
$$H = \int dx\,\, ( N^a \L_a + i \lambda^a \S_a ), \eqno(3.8)$$
vanishes due to the general covariance.

It is easy to see that the action
$$ S = \int d^2x \biggl( b {\dot \psi} + P {\dot \chi} + P_i{\dot f}_i -
{i\over 2} \Gamma_a {\dot \Gamma}_a - {i\over 2}\Theta_a {\dot \Theta}_a -
{i\over 2}\Xi_{a,i} {\dot \Xi}_{a,i}
- N^a \L_a - i\lambda^a \S_a \biggr),
\eqno(3.9)$$
(sum in $i$ and $a$)\hfill \break
is invariant under reparametrization and local supersymmetry, generated
by $\L_a$ and $\S_a$ respectively.
\hfill
\vskip 0.50cm
\centerline{\bf 4. Equations of Motion and Their Solutions for}
\centerline{\bf $2D$ Dilaton (Super)Gravity}
\vskip 0.25cm

In this section we will analyze the equation of motion and their solution
for $2D$ dilaton (super)gravity coupled to conformal matter. This a subtle
problem that was originally studied in the context of $4D$ general
relativity by Dirac {\bf [22]}, de Witt {\bf [23]} and more recently by
Regge and Teitelboim {\bf [24]}. As was emphasized by these authors
the hamiltonian of $4D$ general relativity exhibits some peculiarities that
are not present in other theories. In fact, it can be shown that if we want to
reproduce correctly the Einstein field equations, it is mandatory to add to the
hamiltonian of general relativity a surface term that asymptotically define
the energy and the momentum of the gravitational field. When this
observation is done, one has a well defined variational principle.

At the quantum level, this observation is very important because defining
the surface
term is equivalent to defining a positive ADM mass and, in consequence, to
having a ground state for quantum gravity.

The analogue of this problem for $2D$ dilaton gravity has been considered by
Park and Strominger {\bf [25]} and more recently by Bilal and Kogan
{\bf [13]} and de Alwis {\bf [14]}. In particular in these last references,
the authors use the Regge-Teitelboim method to
obtain an explicit formula
for the ADM mass independent of the time and manifestly positive. Our aim in
this section is to use the Regge-Teitelboim method to obtain the equations
of motion, and we will also write an explicit expression for the surface term,
although as we comment below, we will not impose boundary conditions on the
fields.

Let us start considering the bosonic dilaton gravity.
Following the reference {\bf [24]}, the variation of the hamiltonian
$$\delta H = \int dx \biggl( {\delta H\over \delta \psi}\delta \psi
+ {\delta H\over \delta b}\delta b + {\delta H\over \delta \chi}\delta \chi
+ {\delta H\over \delta P}\delta P + {\delta H\over \delta f_i}\delta f_i
+ {\delta H\over \delta P_i}\delta P_i \biggr), \eqno(4.1)$$
contains a surface term ${\cal D}$ that does not allow one to obtain correctly
the equation of motion. For this reason, it is necessary to redefine the
hamiltonian in the form
$$H \rightarrow H - {\cal D}, \eqno(4.2)$$
in order to cancel the undesired term that appears in (4.1).

By explicit calculation of (4.1), it is found that the surface term is
$$\eqalignno{ {\cal D} = \sum_{a = \pm 1} \biggl[ &
a N^a \delta h_a + a(N^a h_a -
a N^{'a} ) \delta \psi \cr &
- a N^a \delta J_{-a} - {a\over 2}( - N^a J_{-a} +
a N^{'a})\delta \chi \biggr]_{x = -\infty}^{x = +\infty}, &(4.3)\cr}$$
where we have assumed that the matter fields $f_i$
vanishes at $x = \pm \infty$.

Inserting (4.3) in (4.2) we obtain a well defined variational principle
and the equation of motion obtained are
$$\eqalignno{ &
{\dot \psi} = \sum_{a = \pm 1} ( N^a h_a - a N^{'a}),
\cr &
{\dot b} = - \sum_{a = \pm 1} \biggl[ -a {(N^a h_a - a N^{'a})}^{'}
+ {1\over 8} \mu^2 N^a e^\pp \biggr],
\cr &
{\dot \chi} = \sum_{a = \pm 1} \biggl[ 2 ( -N^a J_{-a} + a N^{'a} ) \biggr],
\cr &
{\dot P} = - \sum_{a = \pm 1} \biggl[ {a\over 2} {(- N^a J_{-a} + a N^{'a}
)}^{'} + {3\over 16}\mu^2 N^a e^{\pp} \biggr], &(4.4)
\cr &
{\dot f_i} = \sum_{a = \pm 1} \, Q_{i,a},
\cr &
{\dot P_i} = \sum_{a = \pm 1} \biggl[ \,a N^a Q_{i,a}\biggr]^{'}. \cr}$$

These equations are complicated to solve in an arbitrary gauge and, for this
reason, is convenient to fix the gauge; let us take the proper-time gauge
$$N_+ = 1 = N_-. \eqno(4.5)$$

The previous equations of motion becomes
$$\eqalignno{ & {\dot \psi} - 2 b = 0,
\cr &
{\dot b} - 2 \psi^{''} + {1\over 4}\mu^2 e^{\pp} = 0,
\cr &
{\dot \chi} + 8 P = 0,
\cr &
{\dot P} + {1\over 2} \chi^{''} + {3\over 8}\mu^2 e^{\pp} = 0, &(4.6)
\cr &
{\dot f_i} - 2 P_i = 0,
\cr &
{\dot P_i} - 2 f_i^{''} = 0, \cr}$$
and after eliminating $b, P$ and $P_i$, we obtain
$$\eqalignno{& {\ddot \psi} - 4 \psi^{''} = 2{\mu^2\over 2} e^{\pp},
\cr &
{\ddot \chi} - 4\chi^{''} = -3\mu^2 e^{\pp}, &(4.7)
\cr &
{\ddot f_i} - 4 f_i^{''} = 0. \cr}$$

It is convenient to write these equations in terms of light light
cone coordinates
$x_\pm = \tau \pm {1\over 2} x$. Using this parametrization (4.7) takes the
form\footnote{$^2$}{The first pair of equations is very similar to the Toda
equation, {\it v.i.z.} $\partial_+ \partial_- \Phi_i = e^{K_{ij}} \Phi_j$ where
$K$ is a Cartan matrix of some simple Lie algebra and $\Phi_i$ is a field with
two components {\bf [26]}.}
$$\eqalignno{ &
4 \partial_+ \partial_- \psi = {\mu^2\over 2} e^{\pp},
\cr &
4 \partial_+ \partial_- \chi = -3 \mu^2 e^{\pp}, &(4.8)
\cr &
4 \partial_+ \partial_- f_i = 0, \cr}$$
from these equations we see that the matter fields are trivially integrable
and their solution can be expressed in terms of two chiral functions
$f_i = f_-(x_-) + f_+(x_+)$, which may be determined imposing appropriate
boundary conditions for the matter fields.
The first couple of equations, by other hand, gives the following
relation between $\psi$ and $\chi$
$$\psi + {1\over 6}\chi = \gamma ( g_+ (x_+) + g_- (x_-) ), \eqno(4.9)$$
where $\gamma$ is a constant and $g_\pm ( x_\pm)$ are two chiral functions
that depends, in this case, on the coordinate system.

Using (4.9), for instance, one can bring the second equation in (4.8)
in the form
$$\partial_+ \partial_- {\tilde \chi} = \mu^2 e^{{\tilde \chi}+ \gamma g (x)},
\eqno(4.10)$$
$({\tilde \chi} = {4\over 3} \chi, \,\,g(x) = g_+ (x_+) + g_- (x_-))$.

In order to solve (4.10) one can take a coordinate system where
$g_\pm (x_\pm) = 0$\footnote{$^3$}{This choice corresponds to
choose Kruskal-Szekeres coordinates, see e.g. {\bf [5]} and de Alwis in
{\bf [6]}} and the equation (4.10) becomes the Liouville equation,
whose general solution is {\bf [27]}
$$e^{{\tilde \chi}} = {8\over \mu^2} {h^{'}(x_+) k^{'} (x_-) \over
{[ h (x_+) - k (x_-) ]}^2}, \eqno(4.11) $$
where $h$ and $k$ are two functions that depend on $x_-$ and $x_+$
respectively and that satisfy the restrictions
$$h^{'} > 0, \,\,\,\,\,\,\,\,\,\, k^{'} > 0, \eqno(4.12)$$
$(h^{'} = {dh\over dx})$.

In the same way, one can solve the equation of motion for $\psi$ and by
(4.9) ($g=0$) obtaining
$$e^{-{\tilde \psi}} = {8\over \mu^2} {h^{'} (x_+) k^{'} (x_-) \over
{[ h (x_+) - k_(x_-)]}^2}, \eqno(4.13)$$
$({\tilde \psi} = 8 \psi)$.

If we choose appropriately the functions $h(x_+)$ and $k(x_-)$ and set
the matter fields to zero, the black hole solution discussed by CGHS in
{\bf [5]} is obtained.

Now, let us discuss the equations of motion and their solutions for $2D$
dilaton supergravity.

The variation of the hamiltonian in this case is
$$\eqalignno{ \delta H = \int dx \biggl[&
{\delta H\over \delta \psi} \delta \psi + {\delta H\over \delta b} \delta b
+ {\delta H \over \delta \chi} \delta \chi + {\delta H \over \delta P}
\delta P +
{\delta H\over \delta f_i}\delta f_i
\cr &
+ {\delta H\over \delta P_i} \delta P_i
+ {\delta H\over \delta \Gamma} \delta \Gamma + {\delta H\over \delta \Theta}
\delta \Theta +
{\delta H\over \delta \Xi} \delta \Xi \biggr] + {\cal D}_s. &(4.14) \cr}$$

Using (4.7) we obtain that the equations of motion
$$\eqalignno{ &{\dot \psi} = \sum_{a= \pm 1} \biggl[ N^a h_a - a N^{'a} +
{i\lambda^a\over \sqrt{2}}\Gamma_a \biggr],
\cr &
{\dot b} = - \sum_{a = \pm 1} \biggl[ - a {( N^a h_a - a N^{'a} +
{i\lambda^a \over \sqrt{2}} \Gamma_a)}^{'} + {1\over 8}\mu^2 N^a e^{\pp} +
{i\lambda^a\over 2} ( \omega \Theta_a + \kappa \Gamma_{-a} ) e^{\kk} \biggr],
\cr &
{\dot \chi} = \sum_{a = \pm 1} 2 \biggl( - N^a J_{-a} + a N^{'a} +
{i \lambda^a\over \sqrt{2}} \Theta_{-a} \biggr), &(4.15)
\cr &
{\dot P} = - \sum_{a = \pm 1} \biggl[ {a\over 2}{ ( -N^a J_{-a} + aN^{'a}
+ {i \lambda^a \over \sqrt{2}}\Theta_{-a} )}^{'} + {3\over 16}\mu^2 N^a e^{\pp}
-
{3\over 4}\lambda^a ( \omega \Theta_a + \kappa \Gamma_{-a} ) e^{\kk} \biggr],
\cr &
{\dot \Gamma}_a = i a N^a \Gamma^{'}_a + i{(aN^a \Gamma_a)}^{'} -
{i\over \sqrt{2}}\lambda^ah_a + i \sqrt{2}\lambda^{'a} + i\kappa \lambda^{-a}
e^{\kk},
\cr &
{\dot \Theta}_{-a} = -iaN^a \Theta^{'}_{-a} - i{(aN^a \Theta_{-a})}^{'}
- {i\over \sqrt{2}}\lambda^a J_{-a} + i\sqrt{2}\lambda^{'a} + i\omega
\lambda^{-a} e^{\kk}, \cr}$$
$$\eqalignno{&
{\dot f}_i = \sum_{a=\pm a} ( N^a Q_{a,i} - {i\lambda^a\over \sqrt{2}}
\Xi_{i,a} ),
\cr &
{\dot P}_i = \sum_{a = \pm 1} [ a N^a Q_{a,i} + i a \lambda^a \Xi_{a,i}]^{'},
\cr &
i{\dot \Xi}_{a,i} = 2iaN^a \Xi^{'}_{a,i} - {i\lambda\over \sqrt{2}}Q_{i,a},
\cr}$$
holds if in the canonical hamiltonian (3.8) the surface term
${\cal D}_s$ is added, i.e. if we redefine $H$ by
$$H_{modified} = H (3.8) - {\cal D}_s, \eqno(4.16)$$
where the surface term is
$$\eqalignno{ {\cal D}_s = \sum_{a = \pm} \biggl[& a N^a \delta h_a +
a ( N^a h_a - a N^{'a} + {i \lambda^a \over \sqrt{2}}\Gamma_{-a}) \delta \psi
- a N^a \delta J_{-a}
\cr &
 - {a\over 2} ( - N^a J_{-a} + a N^{'a} +
{i \lambda^a\over \sqrt{2}} ) \delta \chi
- i a N^a \delta \Gamma_a \Gamma_a - i \sqrt{2} \delta \Gamma_a \lambda^a
\cr &
+ i N^a \delta \Theta_{-a} \Theta_{-a} + i \sqrt{2} \lambda^a \delta
\Theta_{-a} \biggr]_{-\infty}^{+\infty}, &(4.17) \cr}$$
(we have supposed, such as in the bosonic case, that the matter fields
vanish for $x = \pm \infty$).

By the same reasons given in the bosonic case, we have choose the proper time
gauge
$$ N_+ = 1 = N_-, \,\,\,\,\, \lambda_+ = \xi = \lambda_-, \eqno(4.18)$$
where $\xi$ is a constant nilpotent spinor.

In this gauge the equations (4.15) take the following form
$$ \eqalignno{ & {\dot \psi} = 2 b + {i\xi\over \sqrt{2}}
( \Gamma_+ + \Gamma_-),
\cr &
{\dot b} = 2 \psi^{''} - {i\xi\over \sqrt{2}} {(\Gamma_+ - \Gamma_-)}^{'} +
{1\over 4} \mu^2 e^{\pp} + {i\xi\over 2}[ \omega (\Theta_+ + \Theta_- )
+ \kappa ( \Gamma_- + \Gamma_+) ] e^{\kk},
\cr &
{\dot \chi} = - 8P + i \sqrt{2} \xi ( \Theta_- + \Theta_+ ), &(4.19)
\cr &
{\dot P} = -{1\over 2} \chi^{''} + {i\xi \over 2\sqrt{2}}{(\Theta_+ -
\Theta_- )}^{'} - {3\over 8} \mu^2 e^{\pp} - {3i\xi\over 4} [ \omega (
\Theta_+ + \Theta_- ) + \kappa ( \Gamma_+ + \Gamma_- ) ] e^{\kk},
\cr &
 i{\dot \Gamma}_a = 2i a \Gamma^{'}_a - {i\over \sqrt{2}}
\xi h_a + i \kappa \xi e^{\kk},
\cr &
i{\dot \Theta}_{-a} = -2ia \Theta^{'}_{-a} - {i\over \sqrt{2}} \xi J_{-a} +
i \omega \xi e^{\kk},
\cr}$$
$$\eqalignno{&
{\dot f}_i = 2 P_i - {i\xi\over \sqrt{2}} (\Xi_{i,+} + \Xi_{i,-}),
\cr &
{\dot P}_i = 2 f_i^{''} + i\xi {(\Xi_{i,+} - \Xi_{i,-})}^{'},
\cr &
i{\dot \Xi}_{a,i} = 2ia \Xi^{'}_{a,i} - {i\xi\over \sqrt{2}}Q_{i,a}.
\cr}$$
After the elimination the momenta $b$ and $P$, these equations are
$$\eqalignno{{\ddot \psi} - 4 \psi^{''} =& {i\xi\over \sqrt{2}}
({\dot \Gamma_+} + {\dot \Gamma_-}) - i\sqrt{2}\xi {(\Gamma_+ - \Gamma_-)}^{'}
+ {\mu^2\over 2} e^{\pp} \cr & + i\xi [ \omega (\Theta_+ + \Theta_-) +
\kappa ( \Gamma_- + \Gamma_+ ) ] e^{\kk}, &(4.19a)\cr&}$$
$$
\eqalignno{ {\ddot \chi} - 4 \chi^{''} = & i\sqrt{2}\xi
({\dot \Theta}_- + {\dot \Theta}_+ ) - i2\sqrt{2} \xi
{(\Theta_+ - \Theta_-)}^{'} - 3\mu^2 e^{\pp}
\cr &
+ 6i \xi [ \omega (\Theta_+ + \Theta_-) + \kappa ( \Gamma_+ + \Gamma_-)
e^{\kk}, &(4.19b) \cr}$$
$$
i{\dot \Gamma}_a - 2ia \Gamma^{'}_a = - {i\over \sqrt{2}} \xi h_a + i\kappa
\xi e^{\kk}, \eqno(4.19c)$$
$$
i{\dot \Theta}_{-a} + 2ia \Theta^{'}_{-a} = -{i\xi\over \sqrt{2}} J_{-a}
+ i \omega \xi e^{\kk}. \eqno(4.19d) $$
$$ {\ddot f}_i - 4 f^{''}_i = i\xi {(\Xi_{i,+} - \Xi_{i,-})}^{'} -
{i\xi\over \sqrt{2}} ( {\dot \Xi}_{i,+} + {\dot \Xi}_{i,-}), \eqno(4.19e)$$
$$ i{\dot \Xi}_{a,i} = 2ia \Xi^{'}_{a,i} - {i \xi\over \sqrt{2}}Q_{i,a}.
\eqno(4.19f)$$

These equations can be greatly simplified using the following observation: the
set of equations (4.19) has the structure
$$D\,X = \xi (\,something\,), \eqno(4.20)$$
where $D$ is a linear operator like $D = a_2 (x){d^2\over dx^2} +
a_1(x){d\over dx} + a_0 (x)$ and $X = ( \psi,\, \chi,\, \Gamma, \, \Theta, \,
\Xi)$. If we multiply (4.20)
by $\xi$ and use the fact that $\xi^2 = 0$, we obtain the equation
$$\xi D\, X = 0. \eqno(4.21)$$

This result has the following implications: i) If the variables $X$ are
bosonic fields, (4.21) implies
$$D\, X = 0, \eqno(4.22)$$
because $\xi$ is a nonzero spinor; ii) If the variables $X$ are
fermionic fields, the complete expression (4.21) must be kept.

Multipling (4.19) by $\xi$ and using light cone like coordinates, we
find that the equations (4.19a), (4.19b) and (4.19e) become
$$\eqalignno{& 4 \partial_+ \partial_- \psi = {\mu^2\over 2} e^{\pp}, &(4.23a)
\cr &
4 \partial_+ \partial_- \chi = -3\mu^2 e^{\pp}, &(4.23b)
\cr &
4\partial_+ \partial_- f_i = 0, &(4.19c) \cr}$$
while (4.19c),(4.19d) and (4.19f) may be written in the
form\footnote{$^4$}{Here
we have used the identity $$\partial_\tau \mp 2a \partial_x =
(1 \pm a)\partial_- + (1 \mp a) \partial_+ = 2 \delta_{\pm, a}\partial_-
+ 2 \delta_{\mp, a} \partial_+.$$}
$$\xi \partial_\mp T_{\pm} = 0, \eqno(4.24)$$
where $T_\pm$ stands for $\Gamma_\pm,\, \Theta_\mp$ and
$\Xi_\pm$.

The equations (4.23) are exactly the equations of motion for
$2D$ dilaton gravity coupled to conformal matter while (4.24)
appears as chirality conditions on the fermionic fields.

When the matter fields $(f, \Xi)$ are put to zero, such in the bosonic case,
we obtain the same black hole solution found in {\bf [5]} plus the chirality
condition (4.24) for $\Gamma$ and $\Theta$. Thus, we conclude that the
super black holes  are the ones of $2D$ dilaton gravity modulo chirality
conditions on the fermionic fields. Exactly the same situation
holds (although the
derivation is more involved) in $4D$ supergravity {\bf [27]}.
\vfill \eject
\hfill
\vskip 0.25cm
\centerline{\bf 5. Conclusions}
\vskip 0.25cm

In this paper we have studied $2D$ dilaton supergravity following canonical
methods. We believe that this point of view is more convenient and permits
one to derive the theory directly in analogy with string theory.
Our procedure is also
an alternative way to derive this theory and as a result, can be made
free of the subtleties associated with supersymmetrization of
Liouville-like theories.

We have also presented a simple derivation of a no-hair theorem that is the
natural analogue of the four dimensional case.

Several problems, such as the explicit form of the ADM mass for $2D$
supergravity and the quantum corrections to effective action associated with
the model presented here are, presently under research.
\vskip 0.15cm
\centerline{\bf Acknowledgments}

We would like to thank M. Ba\~nados, M. Dubois-Violette, C. Teitelboim and
J. Zanelli for discussions. We thank also Norman Fuchs for reading the
manuscript.
\vskip 0.5cm
\centerline{\bf References}
\vskip 0.10cm
\item{{\bf [1]}} For a review on Quantum Gravity see e.g. E. Alvarez,
{\it Rev. Mod. Phys.} {\bf 61}(1989)561.
\item{{\bf [2]}} G. t' Hooft, {\it Black Holes and the Foundations of
Quantum Mechanics}, 1987-85-0040 (Utrecht) preprint (unpublished).
\item{{\bf [3]}} S.W. Hawking, \CMP $\,\,${\bf 43}(1975)199.
\item{{\bf [4]}} S.W. Hawking, \PR $,\,${\bf D11}(1976)2460.
\item{{\bf [5]}} C.G. Callan, S. Gidding, J. Harvey and A. Strominger,
\PR $\,\,${{\bf D45}}(1992)R1005.
\item{{\bf [6]}} J. Russo, L. Susskind and L. Thorlacius, \PL
{\bf B292}(1992)13; S. de Alwis, \PL$\,\,$ {\bf B289}(1992)278; S. de Alwis,
\PR$\,\,$ {\bf D46}(1992)5431; A. Bilal and C.G. Callan, {\it Liouville
Models of Black Holes Evaporation}, PUPT-1320 preprint; T. Banks, A. Dabholkar,
M. Douglas and M. O'Loughin, \PR$\,\,$ {\bf D45}(1992)3607; B. Birnir,
S. Giddings, J. Harvey and A. Strominger, \PR$\,\,$ {\bf D46}(1992)638;
S. Giddings and A. Strominger, {\it Quantum Theories of Dilaton Gravity},
UCSB-TH-92-28 preprint; S. Hirano, Y. Kazama and Y. Satoh, {\it Exact Operator
Quantization of a Model of Two Dimensional Dilaton Gravity} UT-Komaba
93-3 preprint;
T. Uchino, {\it Canonical Theory of 2D Gravity Coupled to Conformal Matter}
TIT/HEP-216-93 preprint; A. Mikovic, {\it Exactly Solvable of 2D Dilaton
Quantum Gravity}, QMW/PH/92/12; F. Belgiorno, A.S. Cattaneo, F. Fucito
and M. Martellini, {\it A Conformal Affine Toda Model of $2D$ Black Holes,
The End-Point State and the $S$-Matrix} Roma preprint; Y. Tanii, \PL
{\bf B302}(1993)191; K.-J. Hamada and A. Tsuchiya, {\it Quantum Gravity and
Black Holes Dynamics in $1+1$ dimensions}; E.J. Martinec and S.L. Shatashvili,
{\it Nucl. Phys.} {\bf B368}(1992)338; E. Witten, {\it Phys. Rev.}{\bf D44}
(1991)314.
\item{{\bf [7]}} O. Lechtenfeld and C. Nappi, \PL $\,\,${\bf B288}(1992)72.
\item{{\bf [8]}} J. Russo and A.A. Tseytlin, {\it Scalar-Tensor Quantum
Gravity in Two Dimensions}, SU-ITP-92-2 preprint.
\item{{\bf [9]}} See e.g. L. Brink and M. Henneaux, {\it Principles of
String Theory}, Plenum Press 1988.
\item{{\bf [10]}} R. Marnelius, {\it Nucl. Phys.} {\bf B211}(1982)14,
{\bf 221}(1982)409.
\item{{\bf [11]}} E.S. Egorian and R.P. Manvelyan,
{\it Mod. Phys. Lett.} {\bf A5}(1990)2371.
\item{{\bf [12]}} E. Abdalla, M.C.B. Abdalla, J. Gamboa and A. Zadra,
\PL $\,\,${\bf B273}(1992)222.
\item{{\bf [13]}} A. Bilal and I.I. Kogan, {\it Hamiltonian Approach
to 2D Dilaton Gravities and the Invariant ADM Mass}, PUPT-1379 preprint.
\item{{\bf [14]}} A. de Alwis, {\it Two Dimensional Quantum Dilaton
Gravity and the Positivity of Energy}, COLO-HEP-309 (1993) preprint.
\item{{\bf [15]}} E. Verlinde and H. Verlinde, {\it A Unitarity $S$-Matrix
for $2D$ Black Holes Formation and Evaporation}, IASS-HEP-93/8 preprint
\item{{\bf [16]}} P.A.M. Dirac, {\it Proc. of Royal Soc.} {\bf 38}(1928)610.
\item{{\bf [17]}} F.A. Berezin and M.S. Marinov,
{\it Ann. Phys. (N.Y.)} {\bf 104}(1977)336.
\item{{\bf [18]}} C. Teitelboim, Phys. Rev. Lett. {\bf 38}(1977)1106.
\item{{\bf [19]}} see e.g. ref. 9
\item{{\bf [20]}} P. van Niewenhuizen, Phys. Rep. {\bf 68}(1981)189.
\item{{\bf [21]}} J. Gamboa and C. Ram\'{\i}rez, \PL$\,\,$ {\bf B301}(1993)20.
\item{{\bf [22]}} P.A.M. Dirac, {\it Proc. of Royal. Soc.} {\bf 114}(1959)924.
\item{{\bf [23]}} B.S. de Witt, \PR $\,\,${\bf 160}(1967)1113
\item{{\bf [24]}} T. Regge and C. Teitelboim, {\it Ann. Phys. (N.Y.)}
{\bf 88}(1974)286.
\item{{\bf [25]}} Y. Park and A. Strominger, \PR$\,\,$ {\bf D47}(1993)1569;
see also,
A. Bilal, {\it Positive Energy Theorem and Supersymmetry in Exactly Solvable
Quantum Corrected $2D$ Dilaton Gravity}, PUPT-1373 (1993) preprint.
\item{{\bf [26]}} For a review on (super)Toda equation see, K. Aoki and
E. D'Hoker, {\it Geometrical Origin of Integrability for Liouville and Toda
Theory}, UCLA/92/TEP/36 preprint.
\item{{\bf [27]}} L. Johanson, A. Kihlberg and R. Marnelius, \PR $\,\,$
{\bf D29}(1984)2798; P. Mansfield, \PR$\,\,$ {\bf D28}(1983)391;
E. D'Hoker, \PR$\,\,$ {\bf 28}(1983)1346.
\item{{\bf [28]}} P.C. Aichelburg and R. G\"uven, \PR$\,\,$
{\bf D24} (1981) 2066,
$\,\,$\PR {\bf D27} (1983) 456; L.F. Urrutia, \PL$\,\,$ {\bf B82} (1979) 52;
P. Cordero and C. Teitelboim, \PL $\,\,${\bf B78} (1978) 80.
\end